# Text classification based on ensemble extreme learning machine

Ming Li *Member, IAENG*, Peilun Xiao, and Ju Zhang, *Member, IAENG*

*Abstract*—In this paper, we propose a novel approach based on cost-sensitive ensemble weighted extreme learning machine; we call this approach AE1-WELM. We apply this approach to text classification. AE1-WELM is an algorithm including balanced and imbalanced multiclassification for text classification. Weighted ELM assigning the different weights to the different samples improves the classification accuracy to a certain extent, but weighted ELM considers the differences between samples in the different categories only and ignores the differences between samples within the same categories. We measure the importance of the documents by the sample information entropy, and generate cost-sensitive matrix and factor based on the document importance, then embed the cost-sensitive weighted ELM into the AdaBoost.M1 framework seamlessly. Vector space model(VSM) text representation produces the high dimensions and sparse features which increase the burden of ELM. To overcome this problem, we develop a text classification framework combining the word vector and AE1-WELM. The experimental results show that our method provides an accurate, reliable and effective solution for text classification.

## I. Introduction

Text classification is to classify text documents into a set of predefined categories. As one of the key technologies of text mining, text classification is widely used in the fields of information retrieval, search engine, question answering system, public opinion analysis and emotional analysis.With the rapid development of network technology, the number of web pages grows exponentially, and efficient and personalized information retrieval needs to develop more accurate and effective text classification technology.At present, the most popular text classification methods include K-NN[1-2], Naive Bayes[3], decision tree[4], Maximum entropy[5-6], support vector machine (SVM) [2,7], neural networks [8-9], fuzzy theory [10] and so on.

Extreme learning machine is a fast-developing technology of machine learning in recent years. ELM is a simple and effective single hidden layer feedforward neural network(SLFNs) learning algorithm[11]. Its model can be obtained analytically which avoids the convergence difficulties existing in conventional methods, so it has a very fast learning speed. SVM has been regarded as one of the most successful methods of traditional text classification methods. A large amount of literature have shown ELM outperformed many of the traditional classifiers including the SVM classifier[14,18,21,22]. ELM also has some shortcomings, and its predictive performance is still affected by the input weights of the neural network and the bias of hidden layer [23]. Due to the input weights of the neural network and the bias of hidden layers random initializing, the performance of ELM is inconsistent when the same experiment is implemented with the condition of the same training and test samples. That means stability of ELM is not ideal. Ensemble learning model can improve classification performance compared with a single model effectively[24]. AdaBoost is an important ensemble learning technology, which could enhance a weak classifier which is slightly better than a random guess to a strong classifier. Yunliang Jiang et al. used PCA to reduce the face features and embedded ELM in the AdaBoost framework for face recognition, experiments obtained good results[25]. Huang haibo et al. proposed ELM algorithm based on AdaBoost to predict the quality of the shock absorber, extracting the abnormal characteristic information of the shock absorber through wavelet packet decomposition. Experiments demonstrated that AdaBoost method could improve ELM performance significantly[26]. Yan Xu et al. applied ELM algorithm to AdaBoost for identification of the traffic signs, also achieved satisfied results[27]. However, these methods did not consider the imbalance of datasets. Kuan Li put forward a weighted ELM based on boosting, which embedded weighted ELM into boosting method and adjusted the distribution weights of samples in each iteration of AdaBoost[29]. However, this method considered only the imbalance between categories and did not consider the imbalance within the categories.

In this paper, we propose a novel approach based on cost-sensitive ensemble weighted extreme learning machine(AE1-WELM) and apply this approach to text classification. We first use the word vector to obtain the document vector of high quality and low dimension, then measure the importance of documents by the category

Manuscript received February 3, 2018; revised July XX, 20XX. This work is supported by Natural Science Foundation of China(No.61672488), Ministry of science and Technology "Key technologies of educational cloud" (No.2013BAH72B01), Chongqing Science & Technology Commission(No.Cstc2015shms-ztzx10005), Clinical Novel Technology Funding of Southwest Hospital(No.SWH2016ZDCX1008 ).

Ming Li is with High performance computing application R&D Center, Chongqing Institute of Green and Intelligent Technology, Chinese Academy of Sciences, Chongqing 400714, China; University of Chinese Academy of Sciences, Beijing100049, China; Center for Speech and Language Technology,Research Institute of Information Technology, Tsinghua University,Beijing,100084,China. (phone: +86-13366712430; fax: 86-023-65935508; e-mail: liming@cigit.ac.cn).

Peilun Xiao is with Center for Speech and Language Technology,Research Institute of Information Technology, Tsinghua University,Beijing,100084,China.; College of Science & Engineering, The University of Edinburgh, Edinburgh EH1,England. (e-mail: Xiaopeilun.bupt@gmail.com).

Ju Zhang is with High performance computing application R&D Center, Chongqing Institute of Green and Intelligent Technology, Chinese Academy of Sciences, Chongqing 400714, China. (e-mail:Zhangju@magicalthink.com).

information entropy, and generate cost-sensitive matrix and factor based on the document importance. We use the cost-sensitive weighted ELM(AE1-WELM) as the base classifier, in which each document weight is adjusted by a cost-sensitive factor in each iteration, then we use each document vector as input for AE1-WELM. Finally, we conduct all the experiments on three standard datasets of 20Newsgroups, Reuters-21578 and WebKB. Experiments show that the proposed method could achieve more balanced results than other ELM methods on imbalance datasets and better results than other ELM methods on balanced datasets.

The rest of the paper is structured as follows: Section 2 introduces ELM related works on text classification and weighted ELM; Section 3 describes the proposed text classification method based on cost-sensitive ensemble WELM in detail; Section 4 presents the experiments and discusses the result; and Section 5 summarizes the conclusion and suggests the future work.

## II. A PREVIEW AND RELATED WORK

### A. Extreme learning machine for text classification

ELM has been applied to text classification in recent years, because of that ELM has faster learning speed and better generalization performance compared with conventional machine learning method. Ying Liu et al. examined the performance of ELM and SVM in text classification[13]. Wenbin Zheng used LSA to reduce the text dimension, then compared the performances of RELM, Neural Networks and SVM in text classification. Experimental results showed RELM had faster-learning speed and better classification performance than other methods[14]; After that, Wenbin Zheng et al. proposed a linear classifier based on non-negative matrix decomposition and a text fast classification framework based on ELM classifier[15]; Xiang guo Zhao et al. proposed an XML document classification framework based on Bagging-ELM. In this framework, they improved the Bagging-ELM algorithm and applied the Revoting of Equal Votes(REV) and Revoting of Confusing Classes (RCC) methods to Bagging-ELM successfully. The improved approach achieved better results than Bagging-ELM[16]. After that, Xiang guo Zhao continued to improve the method, they introduced ε parameter into the RCC method and calculated the probability of the voting results, all of these methods improved the classification performance further[17]. Li juan Duan used KELM to classify historical patent documents and achieved better results than SVM[18]. Yu haiyan et al. reduced the text feature by information gain and introduced wavelet into KELM to conduct the emotional classification of Chinese text[19]. Li Yongqiang proposed CPSO-ELM algorithm to select the input weights of neural network and biases of hidden nodes in ELM to classify XML documents by optimizing search strategy[20]. Rajendra Kumar Roul studied the enhancement of ELM classification performance under feature extraction, did a larger amount of experiments on single ELM and multi-layer ELM which exceeded many state-of-the-art methods, including SVM method[21]. After that, Rajendra Kumar Roul proposed a text feature selection algorithm based on k-means combined with Wordnet to reduce the text dimension, which was collaborated with single ELM and multi-layer ELM for text classification[22].

### B. Weight extreme learning machine

Weighted ELM introduces a weighted matrix W into ELM, which assigns different samples different weight. WELM alleviates imbalance between samples in different categories, thus improves the overall prediction accuracy[33].

Given a set D contains arbitrary distinct $N$ samples $D = \{(x_i, y_i) \mid x_i \in R^n, y_i \in R^m, i = 1, 2, \ldots, N\}$, where $x_i$ is the sample, $y_i$ is the class label, $g(x)$ is the hidden layer activation function, ELM mathematical model can be expressed as:

$$Minimize: \frac{1}{2}\beta^2 + C\frac{W}{2}\sum_{i=1}^{N}\varepsilon^2$$
$$Subject\ to: \beta_i g(ax_i + b_i) - y_i = \varepsilon_i \quad (1)$$
$$i = 1, \ldots, N$$

Derived by KKT conditions:

$$\beta = H^+T = \begin{cases} \left(\frac{I}{C} + H^TWH\right)^{-1} H^TT, & N \geq L \\ H^T\left(\frac{I}{C} + WHH^T\right)^{-1} T, & N < L \end{cases} \quad (2)$$

where $W$ is the diagonal matrix, in which each element value of the diagonal is the weight of each sample. Zong et al. presented two weighting schemes empirically[33]. Given a test sample $x$, the output of WELM is $f(x) = h(x)\beta$. The label of $x$ can be achieved by: $label(x) = \arg\max_i f_i(x), i \in \{1, \ldots, m\}$.

WELM assigns weight to each sample according to the size of the majority category and minority category which the sample belongs to simply. The sample which belongs to the minority category obtain greater weight than that of the sample which belongs to majority category[33]. But all these weights which are assigned to the samples in the same minority category are same, which ignores the imbalances within the same categories, and the samples in the majority categories with the same problem. Li et al. noticed WELM only focused the imbalances between the different categories and improved WELM on this issue. He used different weight-updating schemes for the samples in different categories in the AdaBoost iteration. But he also did not consider the differences of weights between samples within the same categories[29]. The weights of samples within the same categories were updated in each iteration of AdaBoost, but weight updating scheme of samples within the same categories remained same. We also implemented Kuan Li's algorithm in our experiments, we record his algorithm as Ada-WELM in this paper.

## III. TEXT CLASSIFICATION BASED ON COST-SENSITIVE ENSEMBLE WELM

### A. Text representation

A large number of training samples and the high dimension are the characteristics of text classification. The high text dimension will increase the computational burden of the extreme learning machine. The traditional method to deal with this problem is to reduce the text dimension with various text representation methods based on the VSM model, which can reduce the interference of noise to text classification and ensure the accuracy of text classification. The researchers use various text representation based on vector space model (VSM) to represent text in text classification, including TFIDF, LSI, LSA, PLSA traditionally[34]. Compared with the traditional methods of reducing text dimension by feature extraction or feature selection, word vector representation has better feature representation ability. Word vector maps each term(the term means distinct word in the text datasets in this paper) to a low dimensional real vector by training the unlabeled corpus and can avoid the dimension disaster of the text feature effectively[31].

Mikolov et al. proposed two word vector learning models: the CBOW(Continuous Bag of Words) and the Skip-gram model[32]. The Skip-gram model takes the current word as input into a logarithmic linear classifier and predicts the words in the context. Given a sequence of terms $w = \{w_1, w_2, \cdots w_N\}$, $N$ is the length of the sequence; the $i$ th term in the input term sequence is $w_i$. The objective function maximized by the Skip-gram model is shown in formula (3), with the context of current term $w_i$ and prediction window size $b$:

$$\frac{1}{N}\sum_{i=1}^{N}\sum_{-b \leq j \leq b, j \neq 0} \log p(w_{i+j} | w_i) \quad (3)$$

where $p(w_{i+j} | w_i)$ is calculated by softmax function is shown in formula (4), which is defined by Skip-gram model:

$$p(w_{i+j} | w_j) = \frac{\exp(c_{w_{i+j}} c_{w_j})}{\sum_w \exp(c_w c_{w_j})} \quad (4)$$

where $w_{i+j}$ and $w_i$ are the word vectors of $w_{i+j}$ an $w_i$. The word vector model used in our proposed approach is Skip-gram model because Skip-gram model has satisfied performance in text classification task than CBOW model. We firstly generate word vectors for each term $w_i = (v_1, v_2, \cdots v_m)$, $m$ represents the dimension of the word vector, we use $c_{i,j}$ to represent the word vector of the $j$ th term in the $i$ th document. Finally, we generate document vectors through formula (5):

$$v_i = (1/J_i)\sum_{j=1}^{J_i} c_{i,j} \quad (5)$$

where $J_i$ represents the number of terms in the $i$ th document.

*B. Sample Category Information Entropy*

The weight of each term can reflect its importance in the document, and the importance of each term can reflect the term classification ability. The terms with the considerable ability to distinguish documents could help to distinguish categories of documents. The more terms with high text classification ability are in each document, the higher text classification ability each document will be. So we use Shannon's information entropy to measure the importance of each term to construct the importance of each document.

We construct two information entropy function of each term; then we combine these two functions to measure the importance of each term and each document. One is to describe each term distribution in all documents of the whole dataset, we call it as inter-class information entropy function. The other is to describe each term distribution in the documents only from the same category, we call it as inner-class information entropy function.

1) Inter-class information entropy

Definition 1: Given a training set $D = \{d_1, d_2, \ldots, d_n\}$ has m categories $c_j(j = 1, 2, \ldots, m)$, the frequency of term $t_i$ ($t_i$ is the $i$ th term in the document) in the document of the category $c_j(j = 1, 2, \ldots, m)$ is $DF_{ij}$, the frequency of the term $t_i$ in all documents is $DF_i$, the inter-class information entropy function of the term $t_i$ which is recorded as $ED(t_i)$, is defined as:

$$ED(t_i) = \ln\left(\frac{1}{E_d(t_i) + \theta}\right) \quad (6)$$

where $E_d(t_i)$ is the inter-class information entropy of the term $t_i$, $E_d(t_i) = -\sum_{k=1}^{m}\left(\frac{DF_{ij}}{DF_i}\right) \times \ln\left(\frac{DF_{ij}}{DF_i}\right)$, $DF_i = \sum_{j=1}^{m} DF_{ij}$.

According to the information entropy theorem, $E_d(t_i)$ is larger if the term $t_i$ is more uniformly distributed in all categories. $E_d(t_i)$ is smaller if the term $t_i$ is more non-uniformly distributed in each category. The entropy $E_d(t_i)$ is the largest, if and only if the term is uniformly distributed in each category. Firstly, we take the reciprocal form of $E_d(t_i)$. To prevent $E_d(t_i)$ from being zero, we add a parameter $\theta$ to the denominator. We find that $1/(E_d(t_i) + \theta)$ is large usually, which leads to the value of inner-class information entropy is overwhelmed(inner-class information entropy will be mentioned later and defined in this section), which leads to that the effect of inner-class information entropy on text classification is weakened greatly. So, we take the logarithm form of $1/(E_d(t_i) + \theta)$. Finally, we obtain the inter-class information entropy function $ED(t_i)$.

2) Inner-class information entropy

Definition2: Given a training set $D = \{d_1, d_2, \ldots, d_n\}$ has m categories $c_j(j = 1, 2, \ldots, m)$, the frequency of $t_i$ in the $k$ th document in the documents of the category $c_j(j = 1, 2, \ldots, m)$ is $TF(t_i, d_{jk})$, the inner-class information entropy function of $t_i$ which is recorded as $EC(t_i)$, is defined as:

$$EC(t_i) = e^{\max_{j\in[1,m]}(E_c(t_i,c_j))} \quad (7)$$

where $E_c(t_i,c_j) = -\sum_{j=1}^{|c_j|} \frac{TF(t_i,d_{jk})}{TF(t_i,c_j)} \ln\left(\frac{TF(t_i,d_{jk})}{TF(t_i,c_j)}\right)$.

$E_c(t_i,c_j)$ is the inner-class information entropy of the term $t_i$ in the category $c_j$. $|c_j|$ indicates the number of documents in the category $c_j$. $d_{jk}$ represents the $k$ th document in the category $c_j$. $TF(t_i,d_{jk})$ indicates the frequency of the term $t_i$ appearing in the $k$ th document $d_{jk}$ in the category $c_j$. $TF(t_i,c_j)$ indicates the total frequency of term $t_i$ appearing in the category $c_j$. In all categories $c_j (j=1,2,\ldots,m)$, we take the maximum value of $E_c(t_i,c_j)$ as the inner-class information entropy of term $t_i$.

Considering that: 1) Some terms with the high classification ability are low-frequency words, and $\max_{j\in[1,m]}(E_c(t_i,c_j))$ of low-frequency word is small, or even almost equal 0; 2) The value of inner-class information entropy function of the low-frequency words is quite small, or even equal 0, if we use $\max_{j\in[1,m]}(E_c(t_i,c_j))$ as the inner-class information entropy function directly. Based on the above two points, we may generate the wrong importance measurement results of each document, if we use $\max_{j\in[1,m]}(E_c(t_i,c_j))$ as the inner-class information entropy function of each low-frequency word, which is then combined with the inner-class information entropy function of each low-frequency word. To deal with above problem, we add e as the adjustment to the inner-class information entropy and obtain the inner-class information entropy function $EC(t_i)$. We observe that $e^{\max_{j\in[1,m]}(E_c(t_i,c_j))}$ used as $EC(t_i)$ has more considerable classification ability than $\max_{j\in[1,m]}(E_c(t_i,c_j))$ used as $EC(t_i)$ directly through our various text classification experiments.

3) Category information entropy

We combine the inter-class information entropy function with the inner-class information entropy function to obtain the category information entropy function of each term:

$$EDC(t_i) = ED(t_i) \times EC(t_i) \quad (8)$$

The category information entropy of each term not only can describe the text distribution between different categories but also can describe the text distribution in the same categories, which has a finer granularity for the description of the text. We use $EDC(t_i)$ of each term to construct category information entropy of each document. The larger the category information entropy of each document is, the higher text classification ability of each document is. The pseudo code of document category information entropy algorithm is described in Algorithm 1:

Step1: Preprocess document $d_i (1 \le i \le n)$ in set $D = \{d_1, d_2, \ldots, d_n\}$;

after preprocessing, each document $d_i$ in set $D = \{d_1, d_2, \ldots, d_n\}$ is represented as a set of terms $d_i = \{t_{i1}, t_{i2}, \ldots, t_{i|d_i|}\}$;

Step2: Calculate the inter-class information entropy function of each term in the training and testing set

   For $i = 1 : |d_n|$ ( $|d_n|$ is the number of terms in $d_i$ document)

     Calculate the $DF_i$ and $DF_{ij}$ of each term in each document;

     Calculate the $ED(t_i)$ of each term in each document according to formula (6);

   end

Step3: Calculate the inner-class information entropy function of each term in training and testing set

   For $j = 1 : m$

     For $k = 1 : |c_j|$ ( $|c_j|$ is the number of documents in category $c_j$ )

       Calculate the $TF(t_i,d_{jk})$, $TF(t_i,c_j)$ and $E_c(t_i,c_j)$ of each term in the each document;

     end

   end

Calculate the $EC(t_i)$ of each term in each document, according to the formula (7);

Step4: Calculate the $EDC(t_i)$ of each term in each document, according to the formula (8);

Step5: Normalize $EDC(t_i)$;

Step6: Calculate the EDC value of each document in the set $D = \{d_1, d_2, \ldots, d_n\} : EDC = \sum_{i=1}^{|d_n|} EDC(t_i)$

*C. AdaBoost cost-sensitive WELM for text classification*

WELM and Boosting-based WELM did not consider the weight difference between samples in the same categories. To deal with this problem, we introduce cost-sensitive into WELM and assign different weights $w_i (i=1,2,\ldots,n)$ to different samples in the same categories and the different categories, then construct the cost-sensitive matrix $W = \begin{bmatrix} w_1 & & \\ & \ddots & \\ & & w_n \end{bmatrix}$ by the text category information entropy. Each element in the diagonal matrix $W$ is $EDC(x_i)$. The cost-sensitive WELM output parameter $\beta$ is made by $\beta = (WH)^+ WT$.

The AdaBoost algorithm combines several weak classifiers and integrates them into be one strong classifier with strong classification ability. Freund and Schapire improved the AdaBoost algorithm which is used for binary classification problems originally. They generated AdaBoost.M1 and AdaBoost.M2 algorithm and used them for multiclassification problems, and also presented the extension form of AdaBoost.M1[30]. We use the extension form of AdaBoost.M1 algorithm in this paper, then embeds WELM into the AdaBoost.M1 framework to generate AEx-WELM algorithm.

The AdaBoost.M1 algorithm adjusts sample distribution by adjusting the sample weight adaptively, which assigns the larger weights to the samples which are misclassified and the smaller weights to the samples which are classified correctly. AdaBoost.M1 distinguishes the importance of samples through weight distribution of samples based on the misclassification rate of samples. The samples which are misclassified are regarded as the important samples in the AdaBoost algorithm. The AdaBoost algorithm assigns a larger weight to these samples. We are inspired by the idea of weight distribution in the AdaBoost iteration, then introduce the cost-sensitive factor into the AdaBoost.M1 framework. We use document category information entropy to characterize the importance of documents and construct the cost-sensitive factor. The weight of each document is updated according to the importance which is made by both of misclassification rate and category information entropy of each document in each iteration. According to the different rules of documents weight updating in our experiments, AEx-WELM algorithm is denoted as AE1-WELM algorithm and AE2-WELM algorithm respectively. AE1-WELM is our proposed approach, in which the weights of documents are updated by formula(9a). And AE2-WELM updates the weights of documents by formula(9b). Ada-WELM updates the weight without $EDC(x_i)$ factor in formula(9) and with the weight updating rules of original AdaBoost algorithm and normalizes the weights of documents according to $D_1(x_i) = 1/\#t_k$ not document category information entropy(see [33] for details). AEx-WELM updates each element $w_i$ in $W \in R^{N \times N}$ in AdaBoost.M1 each iteration. The pseudo code of AEx-WELM for text classification algorithm is described in Algorithm 2:

Step1: Preprocess the training samples and testing documents, remove the stop words and the special symbols, each document is represented as a collection $d_i = \{t_{i1}, t_{i2}, \ldots, t_{i|d_i|}\}$

Step2: Generate the word vector for each term in each document with Word2vec;

Step3: Generate the document vector for each document according to formula(5);

Step4: Generate the category information entropy of each document with Algorithm 1;

Step5: Generate the cost-sensitive matrix $W_t = diag(EDC(x_i))$

Step6: Train the WELM with weight $W_t$ as the weak classifier $h_t(x)$

Step7: Normalize the weight $D_1(x_i) = D_1(x_i) / \sum_{i=1}^{n} D_1(x_i)$

Step8: For $t = 1 : T$ (T is the number of weak classifiers)
  Compute the error of $h_t(x)$:
  $$\varepsilon_m = \sum_{i=1}^{N} D_m(x_i) I(h_m(x_i) \neq y_i)$$
  While $\varepsilon_m \geq 0.01$ and $\varepsilon_m \leq 0.5$
  Updating $\alpha_t$:
  $$\alpha_m = \log \frac{1-\varepsilon_m}{\varepsilon_m} + \log(k-1)$$
  Update the weight of each document, according to the formula(9):
  $$D_{t+1}(x_i) = D_t(x_i) \times \begin{cases} e^{EDC(x_i)\alpha_t} & (9a) \\ EDC(x_i)e^{\alpha_t} & (9b) \end{cases}$$
  Normalize the $D_{t+1}(x_i)$:
  $$D_{m+1}(x_i) = \frac{D_m(x_i)\exp(-\alpha_m I(h_m(x_i) \neq y_i))}{Z_m},$$
  $Z_m = \sum_{i=1}^{N} D_{m+1}(x_i)$ ($Z_m$ is the normalization factor);
  end
end

Step9: Output the class label:
$$\Theta(x) = \arg\max_k \sum_{t=1}^{T} \alpha_t [h_t(x) = k]$$

IV. EXPERIMENTS AND ANALYSIS

A. Database

We implement all methods on three standard datasets. The 20Newsgroups dataset is collected by Ken Lang1, The WebKB dataset is collected by the CMU project2 and the Reuters-21578 dataset is published by DavidD Lewis3. 20Newsgroups is the balanced dataset, and the latter two are imbalanced datasets. The 20Newsgroups dataset contains 20 different categories of English news, which contains a total of 18846 documents. To improve the reliability, all repeat documents and some news heads are removed, which left 11293 and 7528 documents to the training data and testing data. The original WebKB dataset contains about 8300 Elish websites which are divided into seven categories, and we chose the four most commonly used categories, including student, faculty, course, and project subsets in this paper. To improve the reliability,

---

1http://www.cs.cmu.edu/afs/cs.cmu.edu/project/theo20/www/data/news20.html
2 http://web.ist.ult.pt/~acardoso
3https://kdd.ics.uci.edu/databases/reuters21578/reuters21578.html

some repeat documents are removed, which left 2756 and 1375 documents to the training set and testing set. In the Reuters-21578 dataset, 52 of the 90 most commonly used classes is called R52 subset. R52 contains 6532 documents for training and 2568 documents for testing.Evaluation measures

### B. Evaluation measures

Precision, Recall, and F1 values are widely used in classification performance evaluation. For category $c_i$, they are calculated as: $P_i = \frac{a_i}{b_i}$, $R_i = \frac{a_i}{d_i}$, $F_{1i} = \frac{2 P_i R_i}{P_i + R_i}$; $b_i$ is the number of documents in the category $c_i$; $a_i$ is the number of documents identified as $c_i$ correctly; $d_i$ is the number of documents belonging to the class $c_i$.

Micro-F1 and macro-F1 are two ways to evaluate the multiclassification results. Micro-F1 is to calculate the classification results of all documents and average the results. Macro-F1 is to calculate the classification results for each category and average the results. The specific definitions are as follows: $Micro - F1 = \frac{1}{m}\sum_{i=1}^{m} F_{1i}$, $Macro - F1 = \frac{2 \times MacroP \times MacroR}{MacroP + MacroR}$, $MicroP = \left(\sum_{i=1}^{m} a_i\right) / \left(\sum_{i=1}^{m} b_i\right)$, $MicroR = \left(\sum_{i=1}^{m} a_i\right) / \left(\sum_{i=1}^{m} d_i\right)$, $MicroP$ and $MicroR$ represent Precision and Recall of the Micro-F1 respectively. $MacroP$ and $MacroR$ represent Precision and Recall of the Macro-F1respectively.Micro-F1 tends to major categories, and macro-F1 tends to minor categories. To measure the overall performance of the classification, we use micro-F1, macro-F1, training time and testing time to evaluate the classification results.

### C. Experimental Setting

We preprocess the datasets, including removing stop words, removing single characters and non-alphabetic symbols, converting uppercase letters to lowercase letters, and stemming back. We use the word vector training tool word2vec provided by Google to conduct word vector model training4, then generate the document vector of W2V. LDA training and learning tools invented by Blei is used to generate document vector of LDA5. Naïve Bayes, KNN and SVM are constructed by the scikit-learn tool6.

All experiments are conducted with 3.6GHz CPU and 4GB Memory. The related ELM algorithms are all implemented in python. We run each experiment 10 times and take their averaged result as the final result. We choose the tanh function as the activation function of the hidden node in all ELM-based methods. We apply five-fold cross validation method in our experiments. The grid search is used to find the best combination of the number of hidden nodes L and regularization parameter c. $\{10^0, 10^{-1}, ..., 10^{-8}\}$ is the search range for c values, $\{100, 200, ...1000\}$ is the search scope of L; the number of weak classifiers T is set to be 20. A boldface in a Table 1-4 means the best performance when the setting is same.

### D. Performance Evaluation

ELM-based methods and SVM comparison: To verify the performance of our proposed approach, we conduct the experiments on imbalanced datasets (R52 and WebKB) and balanced dataset(20Newsgroups). From Table 1(the results for all methods from Table 1, under the situation when the text dimension is 100), we can see that micro-F1 and macro-F1 performance of AE1-WELM are higher than all the other methods in most cases, except for that Bagging-ELM is superior to AE1-WELM on WebKB. The results for all ELM-based methods are obtained in Figures 1(a)-1(c), under the situation when the (c, L) values are the best (c, L) values for each ELM-based method (except for ELM and Bagging-ELM in which there is hidden node parameter only). From Figures 1(a)-1(c), we see that AE1-WELM is superior to all other ELM-based methods in most cases, which shows that it is effective to combine the cost-sensitive weighted ELM with the AdaBoost framework. We also observe that the other five ELM-based methods perform poor and AE1-WELM method still maintains satisfied performance, when the dimension is relatively low on all datasets. AE1-WELM performs consistently on all datasets, which illustrates the good generalization performance of our approach.

AE1-WELM comparison with ELM and RELM: We observe that ELM performance is worst in all experiments. MF1 and mF1 performance of AE1-WELM are considerably better than that of ELM and RELM on all datasets, especially on 20Newsgroups(balanced dataset) and R52(imbalanced dataset), where there are relatively more documents.

AE1-WELM comparison with Bagging-ELM: From Figures 1(a)-1(c), we observe that the performance of AE1-WELM is slightly higher than that of Bagging-ELM in this paper. See [16] for details) on 20Newsgroups(balanced dataset), overwhelming higher than that of Bagging-ELM on R52(imbalanced dataset). But The performance of Bagging-ELM is slightly better than AE1-WELM when the number of the hidden nodes is larger whatever the document dimension on WebKB(imbalanced dataset). We analyze this result as that WebKB dataset is small dataset and we only take four subsets which are commonly used in WebKB in which there are relatively few documents. On the other hand, we utilize the diversity of text features effectively through random sampling algorithm which leads to improving the text classification performance by the Bagging-ELM method. Even so, AE1-WELM is superior to Bagging-ELM when the number of hidden nodes is small, and MF1 of AE1-WELM is better than Bagging-ELM all the time on WebKB.

TABLE I
COMPARISION OF CLASSIFICATION RESULTS ON
20NEWSGROUPS, REUTERS AND WEBKB

| Method | 20Newsgroups | | Reuters52 | | WebKB | |
|---|---|---|---|---|---|---|
| | mf1 | MF1 | mf1 | MF1 | mf1 | MF1 |
| Naïve Bayes | 0.609 | 0.611 | 0.858 | 0.538 | 0.727 | 0.711 |
| KNN | 0.706 | 0.696 | 0.891 | 0.440 | 0.798 | 0.787 |
| SVM | 0.766 | 0.755 | 0.922 | 0.592 | 0.872 | 0.856 |
| ELM | 0.784 | 0.770 | 0.922 | 0.621 | 0.866 | 0.853 |
| RELM | 0.771 | 0.755 | 0.921 | 0.545 | 0.879 | 0.863 |
| Bagging-ELM | 0.800 | 0.786 | 0.925 | 0.616 | **0.896** | **0.884** |
| Ada-WEM | 0.792 | 0.783 | 0.925 | 0.650 | 0.883 | 0.869 |
| AE1-WEM | **0.804** | **0.794** | **0.938** | **0.682** | 0.892 | **0.884** |
| AE1-WEM | 0.791 | 0.783 | 0.926 | 0.661 | 0.881 | 0.867 |

---

[4] https://code.google.com/p/word2vec
[5] http://www.cs.princeton.edu/~blei/lda-c
[6] http://www.cs.princeton.edu/~blei/lda-c

TABLE II
COMPARISION OF CATEGORIZATION RESULTS ON 20NEWSGROUPS

| Dim | Number of hidden node | Evaluation measures | ELM | RELM | Bagging-ELM | Ada-WLEM | AE1-WELM | AE2-WELM |
|---|---|---|---|---|---|---|---|---|
| 50 | 100 | mf1 | 0.706 | 0.725 | 0.735 | 0.703 | **0.745** | 0.710 |
| | | MF1 | 0.682 | 0.701 | 0.710 | 0.682 | **0.731** | 0.685 |
| | 400 | mf1 | 0.744 | 0.743 | 0.758 | 0.731 | **0.764** | 0.732 |
| | | MF1 | 0.728 | 0.727 | 0.741 | 0.726 | **0.754** | 0.727 |
| | 800 | mf1 | 0.756 | 0.754 | 0.768 | 0.747 | **0.772** | 0.749 |
| | | MF1 | 0.743 | 0.742 | 0.754 | 0.741 | **0.764** | 0.743 |
| 300 | 100 | mf1 | 0.691 | 0.727 | 0.756 | 0.712 | **0.771** | 0.712 |
| | | MF1 | 0.663 | 0.701 | 0.728 | 0.707 | **0.758** | 0.701 |
| | 400 | mf1 | 0.769 | 0.782 | 0.789 | 0.775 | **0.796** | 0.774 |
| | | MF1 | 0.752 | 0.765 | 0.772 | 0.766 | **0.786** | 0.765 |
| | 800 | mf1 | 0.785 | 0.784 | 0.802 | 0.791 | **0.804** | 0.792 |
| | | MF1 | 0.771 | 0.769 | 0.787 | 0.782 | **0.795** | 0.784 |
| 500 | 100 | mf1 | 0.682 | 0.727 | 0.755 | 0.712 | **0.771** | 0.703 |
| | | MF1 | 0.656 | 0.700 | 0.727 | 0.706 | **0.758** | 0.691 |
| | 400 | mf1 | 0.769 | 0.781 | 0.792 | 0.778 | **0.801** | 0.777 |
| | | MF1 | 0.751 | 0.764 | 0.774 | 0.769 | **0.791** | 0.768 |
| | 800 | mf1 | 0.786 | 0.782 | 0.805 | 0.799 | **0.808** | 0.795 |
| | | MF1 | 0.773 | 0.768 | 0.790 | 0.789 | **0.799** | 0.786 |

TABLE II
COMPARISION OF CATEGORIZATION RESULTS ON R52

| Dim | Number of hidden node | Evaluation measures | ELM | RELM | Bagging-ELM | Ada-WLEM | AE1-WELM | AE2-WELM |
|---|---|---|---|---|---|---|---|---|
| 50 | 100 | mf1 | 0.857 | 0.8623 | 0.867 | **0.899** | 0.889 | 0.899 |
| | | MF1 | 0.310 | 0.266 | 0.321 | **0.579** | 0.401 | 0.560 |
| | 400 | mf1 | 0.904 | 0.902 | 0.908 | 0.919 | **0.923** | 0.915 |
| | | MF1 | 0.509 | 0.440 | 0.525 | **0.629** | 0.592 | 0.604 |
| | 800 | mf1 | 0.917 | 0.918 | 0.919 | 0.920 | **0.931** | 0.920 |
| | | MF1 | 0.607 | 0.544 | 0.614 | 0.644 | **0.663** | 0.620 |
| 100 | 100 | mf1 | 0.861 | 0.868 | 0.867 | **0.907** | 0.891 | 0.908 |
| | | MF1 | 0.312 | 0.317 | 0.321 | **0.594** | 0.409 | 0.613 |
| | 400 | mf1 | 0.905 | 0.911 | 0.910 | 0.920 | **0.929** | 0.921 |
| | | MF1 | 0.518 | 0.459 | 0.537 | **0.637** | 0.620 | 0.661 |
| | 800 | mf1 | 0.918 | 0.921 | 0.923 | 0.927 | **0.938** | 0.925 |
| | | MF1 | 0.602 | 0.568 | 0.615 | 0.661 | **0.684** | 0.666 |
| 300 | 100 | mf1 | 0.861 | 0.870 | 0.868 | **0.903** | 0.891 | 0.907 |
| | | MF1 | 0.307 | 0.303 | 0.324 | **0.608** | 0.404 | 0.608 |
| | 400 | mf1 | 0.907 | 0.909 | 0.911 | 0.919 | **0.930** | 0.919 |
| | | MF1 | 0.529 | 0.440 | 0.549 | 0.608 | **0.614** | 0.618 |
| | 800 | mf1 | 0.923 | 0.918 | 0.925 | 0.926 | **0.938** | 0.926 |
| | | MF1 | 0.618 | 0.541 | 0.626 | 0.656 | **0.674** | 0.662 |

TABLE II
COMPARISION OF CATEGORIZATION RESULTS ON WEBKB

| Dim | Number of hidden node | Evaluation measures | ELM | RELM | Bagging-ELM | Ada-WLEM | AE1-WELM | AE2-WELM |
|---|---|---|---|---|---|---|---|---|
| 100 | 100 | mf1 | 0.836 | 0.857 | 0.854 | 0.835 | **0.869** | 0.837 |
| | | MF1 | 0.814 | 0.835 | 0.835 | 0.822 | **0.850** | 0.820 |
| | 400 | mf1 | 0.867 | 0.886 | **0.884** | 0.870 | 0.883 | 0.868 |
| | | MF1 | 0.851 | 0.873 | 0.870 | 0.856 | **0.871** | 0.855 |
| | 800 | mf1 | 0.866 | 0.871 | **0.888** | 0.881 | 0.886 | 0.877 |
| | | MF1 | 0.853 | 0.860 | 0.876 | 0.867 | **0.879** | 0.863 |
| 200 | 100 | mf1 | 0.834 | 0.857 | 0.854 | 0.837 | **0.867** | 0.832 |
| | | MF1 | 0.810 | 0.835 | 0.834 | 0.822 | **0.847** | 0.821 |
| | 400 | mf1 | 0.868 | 0.886 | 0.887 | 0.873 | **0.891** | 0.878 |
| | | MF1 | 0.852 | 0.873 | 0.874 | 0.860 | **0.880** | 0.864 |
| | 800 | mf1 | 0.866 | 0.871 | **0.896** | 0.883 | 0.892 | 0.881 |
| | | MF1 | 0.853 | 0.860 | **0.885** | 0.869 | 0.884 | 0.867 |
| 300 | 100 | mf1 | 0.836 | 0.858 | 0.855 | 0.846 | **0.867** | 0.835 |
| | | MF1 | 0.814 | 0.838 | 0.835 | 0.832 | **0.849** | 0.822 |
| | 400 | mf1 | 0.871 | 0.879 | 0.888 | 0.875 | **0.890** | 0.876 |
| | | MF1 | 0.857 | 0.867 | 0.874 | 0.861 | **0.879** | 0.862 |
| | 800 | mf1 | 0.866 | 0.871 | **0.897** | 0.884 | 0.893 | 0.884 |
| | | MF1 | 0.854 | 0.859 | **0.886** | 0.871 | 0.884 | 0.871 |
| | 100 | mf1 | 0.836 | 0.852 | 0.855 | 0.837 | **0.870** | 0.837 |

| | | | | | | | |
|---|---|---|---|---|---|---|---|
| | | MF1 | 0.814 | 0.831 | 0.834 | 0.825 | **0.850** | 0.824 |
| 400 | 400 | mf1 | 0.870 | 0.875 | 0.886 | 0.877 | **0.889** | 0.873 |
| | | MF1 | 0.854 | 0.859 | 0.871 | 0.863 | **0.877** | 0.856 |
| | 800 | mf1 | 0.865 | 0.878 | **0.894** | 0.884 | 0.893 | 0.880 |
| | | MF1 | 0.851 | 0.866 | 0.882 | 0.871 | **0.883** | 0.865 |
| | 100 | mf1 | 0.843 | 0.846 | 0.853 | 0.841 | **0.871** | 0.839 |
| | | MF1 | 0.821 | 0.853 | 0.833 | 0.829 | **0.852** | 0.824 |
| 500 | 400 | mf1 | 0.870 | 0.871 | 0.886 | 0.879 | **0.888** | 0.871 |
| | | MF1 | 0.856 | 0.857 | 0.872 | 0.865 | **0.878** | 0.857 |
| | 800 | mf1 | 0.868 | 0.878 | **0.893** | 0.882 | 0.890 | 0.881 |
| | | MF1 | 0.856 | 0.867 | 0.881 | 0.868 | 0.882 | 0.867 |

AE1-WELM comparison with Ada-WELM and AE2-WELM: The performance of AE1-WELM was significantly higher than the performance of Ada-WELM and AE2-WELM on the imbalanced datasets R52 and WebKB, which indicates that the proposed method can improve the classification performance of imbalance multiclassification problems. The performance improvement of AE1-WELM is also more obvious than Ada-WELM and AE2-WELM on 20Newsgroups(balanced dataset). In AEx-WELM, AE2-WELM performs not as good as AE1-WELM, almost as same as Ada-WELM. The distribution weight based on formula(9b) varies drastically during each iteration in AdaBoot.M1, which leads to AE2-WELM classification performance is not superior. Ada-WELM shows advantage comparison with AE1-WELM when the number of hidden nodes is small, but AE1-WELM is still superior to Ada-WELM for the point of the whole view on WebKB.

We also observe that text representation based on word vector not only reduces the text dimension effectively, but also help the classifiers achieve as good performance in low dimensional space (usually 100 dimensions, even lower dimensions) as the performance of the traditional VSM model in 1000 dimensions(even higher dimensions) from Table 1-4. From Figures 1(a)-1(c), we see that the performance of six ELM-based methods decrease when

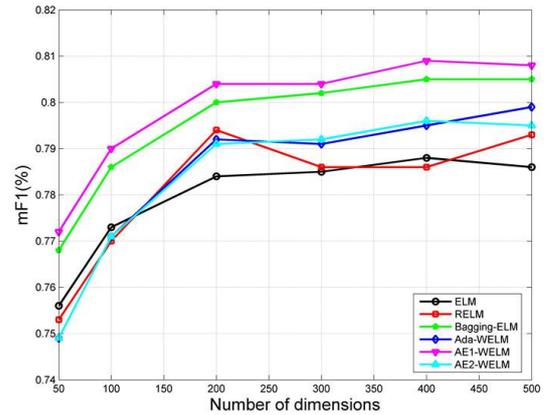
(a)

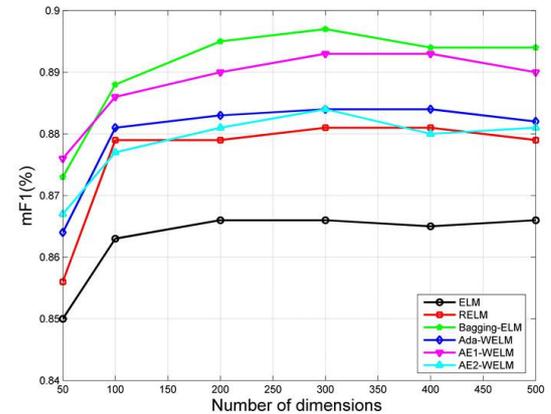
(b)

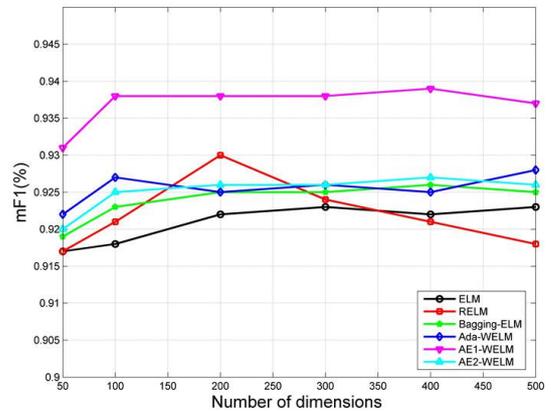
(c)

Fig. 1. Performance on standard datasets.
(a) on 20Newsgroups; (b) on R52 ; (c) on WebKB

document dimension is more than 400, which indicates that the excessively high dimension not only imposes burdens on the ELM but also increases the noise, and decrease the classification performance.

Figures 2(a)-2(c) show mF1 and MF1 performance of AE1-WELM when the number of hidden nodes varies and document dimension is 100 on all datasets. From Figures

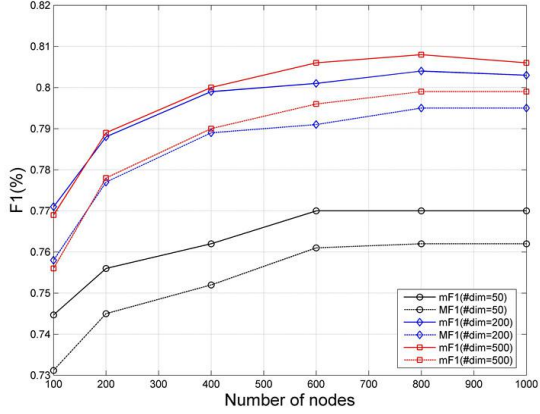

(a)

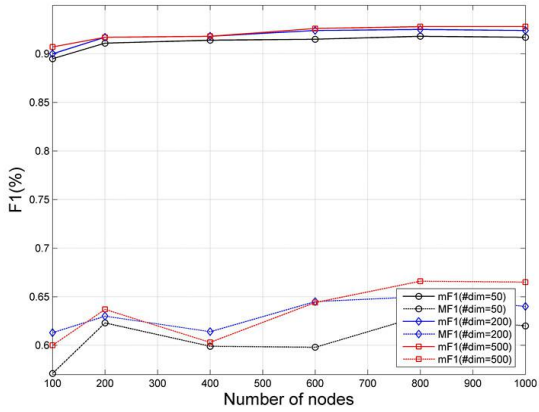

(b)

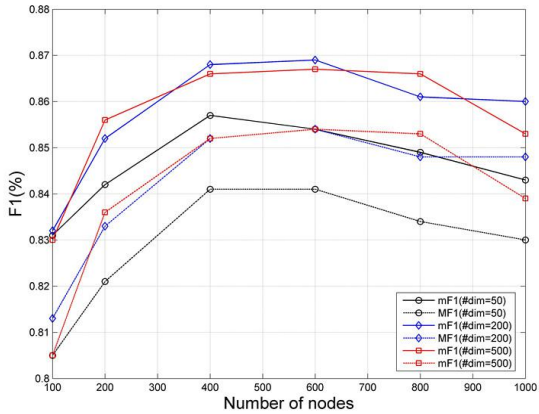

(c)

Fig. 2. Performance of AE1-WELM on standard datasets while #node varies. (a) 20Newsgroups; (b) R52 ; (c) WebKB

2(a)-2(c), we observe that although more hidden nodes can help AE1-WELM achieve better classification results, the mF1 and MF1 performance will reach stability when the number of hidden nodes reaches 600 on 20Newsgroups, 800 on R52, 400 on WebKB. The number of hidden nodes is too large to help improve the classification performance. When the number of hidden nodes exceeds a certain number, the classification performance becomes unstable. When the number of hidden nodes exceeds 800 on 20Newsgroups and R52, 600 on WebKB, the classification performance decreases with the increase of hidden nodes, which should be caused by over-fitting.

From Figures 3(a)-3(c), we observe that the regularization parameter c is a key factor and the classification performance reaches stability with the decrease of c (For c value search range, the empirical suggestion is $\{10^0, 10^{-1},..., 10^{-8}\}$). The effect of regularized parameter c on performance is greater than that of hidden nodes. The performance of AE1-WELM is not sensitive to the selection of hidden node parameters when c decrease to $10^{-5}$; the performance will maintain stable when c dropped below $10^{-5}$ on 20Newsgoups. And the same situation on R52, the difference is that the inflection point is not $10^{-5}$ but $10^{-6}$. We have the same observation on WebKB, and the inflection point becomes $10^{-4}$. There is a difference with 20Newsgroups and R52 is that the performance decreased obviously when the c dropped below $10^{-4}$, and the hidden node number is relatively large.

## V. CONCLUSION

In this paper, we propose a novel ensemble WELM model based on cost-sensitive. Moreover, we propose a text classification framework combining the word vector and AE1-WELM. And we develop an algorithm including balanced and imbalanced multiclassification for text classification. AE1-WELM updates the document distribution through the cost-sensitive factor which is made by document category information entropy in the multiclassification AdaBoost.M1 framework. Experiments show that our method has more competitive classification performance, stability and generalization than SVM and other ELM methods.

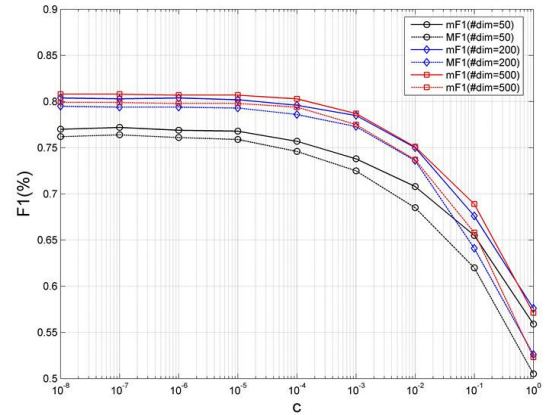

(a)

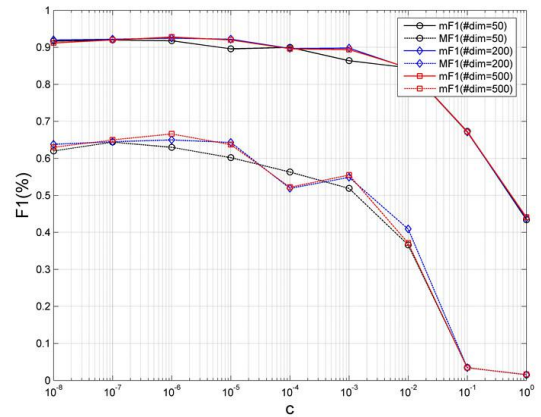

(b)

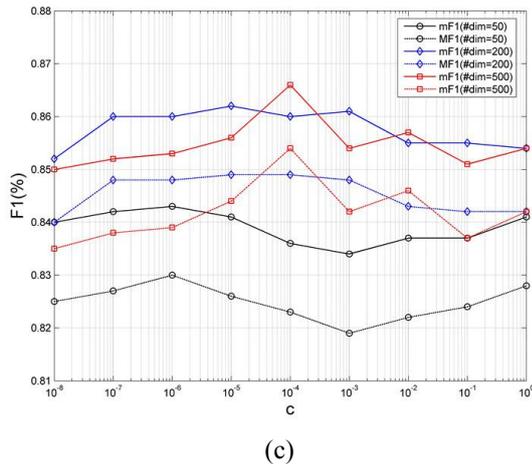

Fig. 3. Performance of AE1-WELM on standard datasets while c varies.
(a) 20Newsgroups; (b) R52 ; (c) WebKB ;

In future, we will study how to reduce text dimension based on word vector model further to construct the better text representation. Further, we will focus on choosing a more reasonable cost-sensitive function to reduce the computational cost of AE1-WELM and optimizing the AE1-WELM framework to obtain better text classification performance.